\documentclass[traditabstract,a4paper]{aa}
\usepackage{txfonts} % A&A font !!

\usepackage{color}

\usepackage{natbib}
\bibpunct{(}{)}{;}{a}{}{,} % to follow the A&A style

\usepackage{graphicx}
\usepackage{fixltx2e}

\usepackage{longtable}
\usepackage{multirow}
\usepackage{overpic}
\usepackage{array}

\usepackage[table,usenames,dvipsnames]{xcolor}
\definecolor{RoyalPurple}{cmyk}{0.75,0.9,0,0.1}
\definecolor{MyBlue}{rgb}{0.0025,0.1125,0.2}
\usepackage[breaklinks, colorlinks, citecolor=blue, linkcolor=MyBlue, urlcolor=RoyalPurple, 
            colorlinks=true, linkcolor=blue, debug, baseurl=' ']{hyperref}

\graphicspath{{Figures/}}

\begin{document}

\title{Bi-layer Kinetic Inductance Detectors for space observations between 80-120 GHz}

\author{A.~Catalano  \inst{1}
\and
J.~Goupy  \inst{2}
\and
H.~le Sueur  \inst{3}
\and
A.~Benoit  \inst{2}
\and
O.~Bourrion  \inst{1}
\and
M.~Calvo \inst{2}
\and
L.~Dumoulin  \inst{3}
\and
F.~Levy-Bertrand  \inst{2}
\and
J.~Mac\'{\i}as-P\'erez  \inst{1}
\and
S.~Marnieros \inst{3}
\and
N.~Ponthieu  \inst{4}
\and
A.~Monfardini  \inst{2}
 }

\offprints{A. Catalano - catalano@lpsc.in2p3.fr} %\email{catalano@lpsc.in2p3.fr} 

\institute{Laboratoire de Physique Subatomique et de Cosmologie, 
  CNRS/IN2P3, Universit\'e Joseph Fourier Grenoble I,
  Institut National Polytechnique de Grenoble, 
  53 rue des Martyrs, 38026 Grenoble Cedex, France 
\and Institut N\'eel F-38042, CNRS, Universit\'e Joseph Fourier Grenoble I F-38042,
  25 rue des Martyrs, Grenoble, France
\and Centre de Sciences Nucl\'eaires et de Sciences de la Mati\`ere (CSNSM), CNRS/IN2P3, bat 104 - 108, 91405 Orsay Campus, France
\and Institut de Plan\'etologie et d'Astrophysique de Grenoble (IPAG), CNRS and Universit\'e de
Grenoble, France
}  

%\date{Received XX, 2013; accepted XX, 2013}

% \abstract
 %{We have developed Lumped Element Kinetic Inductance Detectors (LEKID) sensitive in the band 80-120~GHz for a low optical background. This development is related to the next generation of millimetre-wave space satellites.
%We take advantage of the superconducting proximity effect to reduce the gap of Aluminium, and make it sensitive to radiation below 100~GHz. We have tested a number of arrays based on multi-layers combinations of Aluminium and Titanium and measured their sensitivity using a dedicated closed-circle 100 mK dilution cryostat and a sky simulator based on a pulse tube cooler. The spectral response has been measured with a Martin-Puplett interferometer in a range up to THz with a frequency resolution of 3 GHz. We demonstrated that Ti-Al bilayers LEKID can have an optical sensitivity of about $1.4$ $10^{-17}$~$W/Hz^{0.5}$ for the best pixel and about $2.2$ $10^{-17}$~$W/Hz^{0.5}$ averaged over the array, for an optical background of roughly 0.4~pW per pixel. This performance is close to a sensitivity of twice the CMB photon noise limit at 100~GHz which drove the design of the Planck HFI instrument and it is the baseline for the new generation of millimetre space missions.} %such as COrE+ mission.}
 
 \abstract
 {We have developed Lumped Element Kinetic Inductance Detectors (LEKID) sensitive in the frequency band from 80 to 120~GHz. In this work, we take advantage of the so-called proximity effect to reduce the superconducting gap of Aluminium, otherwise strongly suppressing the LEKID response for frequencies smaller than 100~GHz. We have designed, produced and optically tested various fully multiplexed arrays based on multi-layers combinations of Aluminium (Al) and Titanium (Ti). Their sensitivities have been measured using a dedicated closed-circle 100 mK dilution cryostat and a sky simulator allowing to reproduce realistic observation conditions. The spectral response has been characterised with a Martin-Puplett interferometer up to THz frequencies, and with a resolution of 3~GHz. We demonstrate that Ti-Al LEKID can reach an optical sensitivity of about $1.4$ $10^{-17}$~$W/Hz^{0.5}$ (best pixel), or $2.2$ $10^{-17}$~$W/Hz^{0.5}$ when averaged over the whole array. The optical background was set to roughly 0.4~pW per pixel, typical for future space observatories in this particular band. The performance is close to a sensitivity of twice the CMB photon noise limit at 100~GHz which drove the design of the Planck HFI instrument. This figure remains the baseline for the next generation of millimetre-wave space satellites.}

% keywords must be selected from the A&A list
%\authorrunning{A. Catalano et al.}
\titlerunning{Bi-layer Kinetic Inductance Detectors for space observations between 80-120 GHz}

\maketitle

\section{Introduction} \label{sec1}

The study of the Cosmic Microwave Background (CMB) temperature and polarisation anisotropies has become a powerful tool for cosmology thanks in particular to the spatial observations of the COBE \citep{cobe},
the WMAP \citep{wmap} and Planck satellites \citep{Planck2013gen}. The Planck mission has provided detailed temperature and polarisation CMB maps \citep{cosmo}, but it was not conceived as the ultimate instrument for CMB polarisation measurements and it will most probably only marginally measure primordial CMB polarisation B-modes. The latter are only sourced by tensor perturbations in the early universe and are predicted by inflation \citep{huwhite}. To improve this situation, new CMB space missions such as CORE+\footnote{\url{http://www.core-mission.org/documents/}\\\url{CoreProposal_Final.pdf}}, PIXIE \citep{kogut}, LiteBIRD \citep{Matsumura2014}, are under study. 

In this context, space observations at about and slightly below 100~GHz have proved to be fundamental for CMB science, as they lie in the frequency range for which Galactic foreground contamination is minimal
both in intensity and polarisation \citep{,2014arXiv1409.2495P,2012APh....36...57F,2011A&A...526A.145F}. The Planck mission\footnote{\url{http://www.esa.int/Planck}} was equipped with 11 HEMT (Low Frequency Instrument \cite{PlanckLFIperf}) and~ 52 ~high impedance bolometers (High Frequency Instrument \cite{Planck2011perf}) observing from 30~GHz to 1~THz. Among those, 8 Polarised Sensitive Bolometers (PSB) with a bandpass centred at 100~GHz ($\Delta \nu / \nu = 0.33$) were used to survey the sky in temperature and polarisation at an angular resolution of about 10 arcmin. This provided HFI with a powerful tool to detect and precisely measure the CMB E-modes but not the primordial B-modes which are expected to be much fainter \citep{cosmo}. To meet the challenge of measuring CMB polarisation B-modes, it is necessary to improve the instrumental sensitivity by at least one order of magnitude with respect to Planck. In terms of Noise Equivalent Power this corresponds going from $10^{-17}$~$WHz^{-1/2}$ to $10^{-18}$~$WHz^{-1/2}$. This can be achieved by increasing the focal plane coverage, using thousands of Background Limited Instrument Performance (BLIP) contiguous pixels.
\\
Kinetic Inductance Detectors (KID) \citep{kid} have now reached a maturity adequate for space instruments. The first demonstration of this maturity was achieved by ground-based experiments and in particular by the New IRAM KID Array (NIKA) instrument. NIKA is a dual-band camera operating with frequency-multiplexed arrays of Lumped Element Kinetic Inductance Detectors (LEKIDs) based on Aluminium films and cooled at 100~mK \citep{monfardini}. 
The NIKA instrument exhibits state-of-the-art sensitivity in the 120-300~GHz band range \citep{catalano}. 

The frequency range below 120~GHz is not accessible using the NIKA Aluminium films due to the superconducting gap cut-off. Therefore we need to adopt lower superconducting gap films. Several authors have investigated Titanium Nitride (TiN) as an alternative to Aluminium KID \citep{swenson, leduc, bueno} in the last years. In parallel we have studied a large number of possible alternative solutions. Among them, we have investigated Nb(x)Si(1-x) as a promising alloy to be used at these frequencies \citep{calvo}. Here we use a different approach: we start from Aluminium which has given the best performances up to now, and lower its gap using the superconducting proximity effect with Titanium. 
% \\
%This study particularly relates to the recently submitted COrE+ (Cosmic Origin Explorer)  proposal to answer to the ESA call for a medium-size mission for a launch in 2025. This proposal intends to built a satellite to measure CMB B modes through a high sensitivity survey of the entire sky. The COrE+ instrument requires polarization sensitive direct detectors able to reach the background photon noise limited with a spectral coverage from 60~GHz to 1~THz. Three main technologies are considered:  Transition Edge Sensors (TES) \citep{tes}, Kinetic Inductance Detectors (KID) or Metal Insulator Sensors (MIS) \citep{mis}. 
\\
%This paper is structured as follows: first we give the basic formula to discuss the testing and some details about the design and the fabrication of the bi-layers arrays. In Sec. \ref{perf}, we describe the experimental setup and the results obtained. 
This paper is structured as follows. In Sec. \ref{desc}, we give the main ingredients of the bi-layer LEKID design and describe its fabrication process; details on the noise and responsivity of a LEKID array are also presented. In Sec. \ref{opti}, we describe the experimental setup. Sec. \ref{perf} presents the measurements and the results obtained. We draw conclusions in Sect. \ref{concl}.

\section{LEKID Detectors}\label{desc}

Here we review the basic ingredients necessary to discuss the design and to evaluate the performances of the Ti-Al bilayer LEKID. 
General KID theory was first establish by \cite{kid}. For a complete review, we suggest \cite{Zmuidzinas} and \cite{doylephd}.
%~\footnote{\url{http://www.astro.cardiff.ac.uk/~spxsmd/Lumped_Element_Kinetic_Inductance_Detectors.pdf}}. 
LEKID is a resonator fabricated from superconducting elements in which absorbed photons can change the Cooper pair density producing a change in both resonant frequency and the quality factor of the resonator. LEKID arrays are based on series of LC resonators of different eigenfrequencies,
coupled to a single 50~$\Omega$ feed-line. 
The term \emph{Lumped Element} describes the fact that the resonator is small compared to the wavelength at the resonance frequency. 
%Thereby one can spatially distinguish an inductor (resp. capacitor) part, where the current (resp. charge) density is roughly constant, which maximises the response of the resonator to Cooper pair density variations.
In addition, Lumped Element devices act directly as the absorber of GHz - THz photons, which greatly simplifies the design and array fabrication. Furthermore, one needs to match their sheet impedance to the incoming photons, which implies using either very thin, or quite resistive layers. In that respect, Ti is preferable to noble metals such as Au, Ag or Cu for tuning the superconducting gap of Aluminium.

%We refer as \emph{Lumped Element} if the device itself can act as an absorber of GHz - THz photons. 

\subsection{Tuning the gap cut-off with proximity effect}\label{tc}
The minimum photon energy that can be absorbed in a superconductor below the critical temperature is equal to twice the superconducting gap $\Delta_0$. For standard \footnote{In this paper we name \emph{standard superconductors} those that follows the superconductivity theory proposed by John Bardeen, Leon Cooper, and John Robert Schrieffer (BCS)} superconductors:

\begin{equation}\label{gap}
\Delta_0 \approx 1.764 \cdot k_BT_c
\end{equation}
where $k_B$ is the Boltzmann constant and $T_c$ is the critical temperature of the transition to the superconducting state. 

%In general, tuning $T_c$ amounts to tuning $\Delta_0$.
Adjusting $T_c$ is therefore important in order to tune the absorption of photons to the suitable frequency. 
The possibility of tuning the transition temperature of a superconductor through the proximity effect with a normal metal (or another superconductor) is a well-known property of superconductivity. It has been successfully exploited in transition edge sensors to operate bolometers at a lower temperature and hence lower their specific heat \citep{irwin_tes_2005}. Here we also exploit the proximity effect, but in this case to lower the energy threshold of detectable photons.

A comprehensive calculation of the $T_c$ of bi-layers is given in \cite{2000NIMPA.444...23M}. More generally, all relevant properties (in particular the density of states) can be derived in the framework of the Usadel theory \citep{usadel}, which describes inhomogeneous superconductors. One needs to solve (most often numerically) two coupled non-linear differential equations describing the diffusion of electronic excitations (single and paired) in an inhomogeneous pair potential, provided boundary conditions at the interfaces. Here we consider that the contact between the two layers is perfect and that the average transmission is unity, which provides the strongest proximity effect. Furthermore, the superconducting coherence length is assumed to be larger than the thickness of the film, and thereby superconducting properties are homogeneous over the thickness.

In that case, an approximate analytical solution can be derived following Cooper's intuitive approach \citep{cooper_PE_1961}, considering the bilayer as a superconductor with an effective pairing potential $N V_{eff}$. As electrons spend a fraction of time $N_1 d_1/(N_1 d_1+N_2 d_2)$ in layer 1, this effective potential writes:

$$
N V_{eff} = \frac{ N_1 V_1(N_1 d_1) + N_2 V_2(N_2 d_2)}{N_1 d_1+N_2 d_2},
$$
where $N_x$ is the electron density, $V_x$ the electron phonon coupling and $d_x$ the thickness of layer $x$. This effective potential is then used to calculate $T_c$ in the standard superconductors gap equation $k_B T_c \simeq 1.13 \hbar \omega_D  \cdot e^{(-1/N V_{eff})}$, where $\omega_D$ is the Debye frequency. This approach is valid only when the vibration properties of the two layers are identical, which is roughly the case for Ti and Al. From this simple formula, the general trend of which is confirmed by more thorough calculations, one notices the great convenience of bi-layers since their critical temperature can be finely tuned by changing the relative thickness of the components.
For the Ti~10~nm/Al~25~nm bi-layers fabricated in this work, the expected $T_c$ is between 850~mK and 920~mK. This uncertainty comes from the fact that we need to take into account an increase of the $T_c$ of Al at 25~nm compared to the bulk value (the discussion of which is not the purpose of the present work). In a standard superconductor, this yields a frequency threshold for photon absorption between 62~GHz and 68~GHz.
A last argument for choosing Ti rather than noble metals comes from its weak solubility in Al, which improves predictability, reproducibility and robustness of the devices to aging.

\begin{figure}
\begin{center}
\includegraphics[width=7cm]{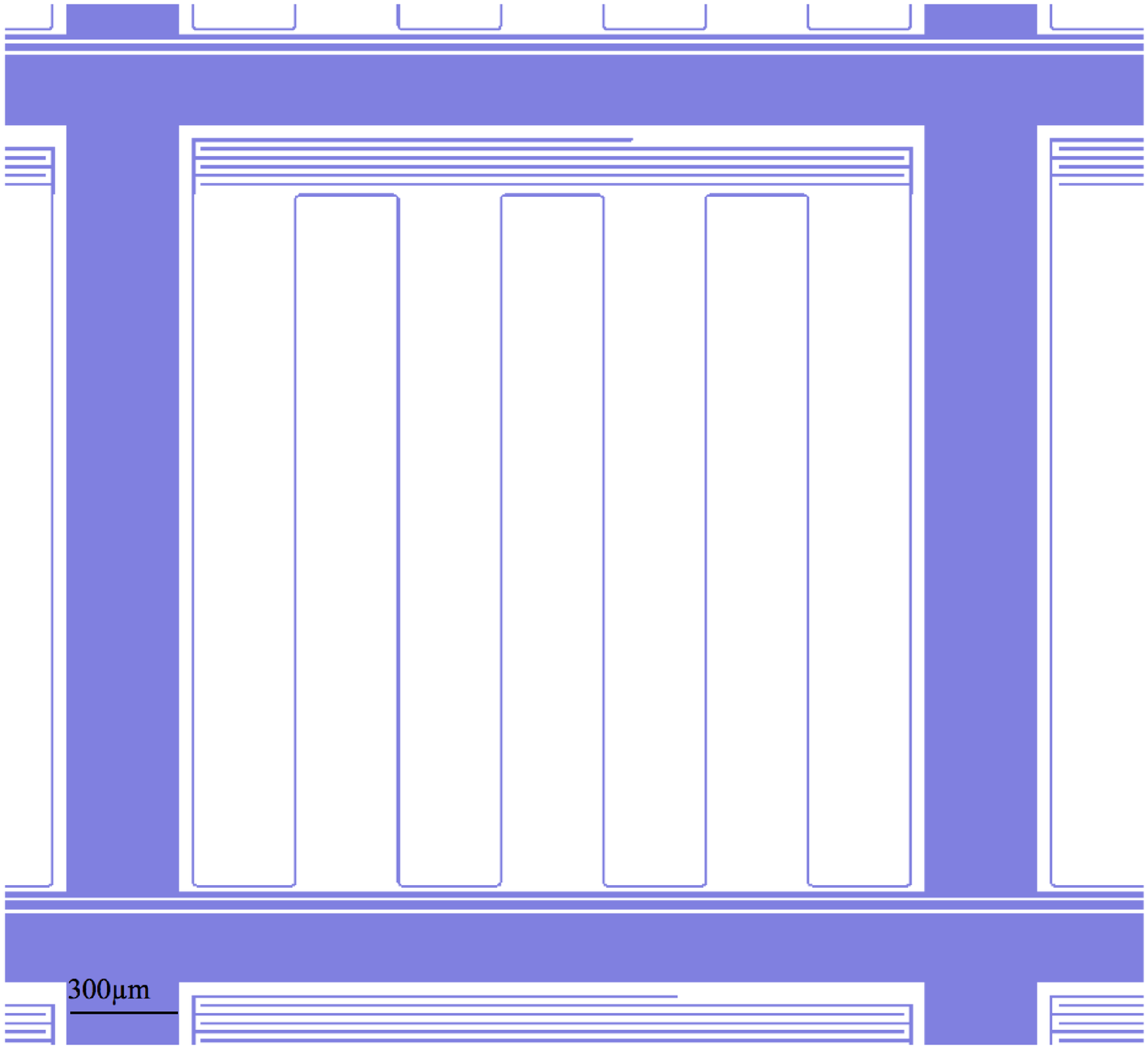}\\
\includegraphics[width=7cm]{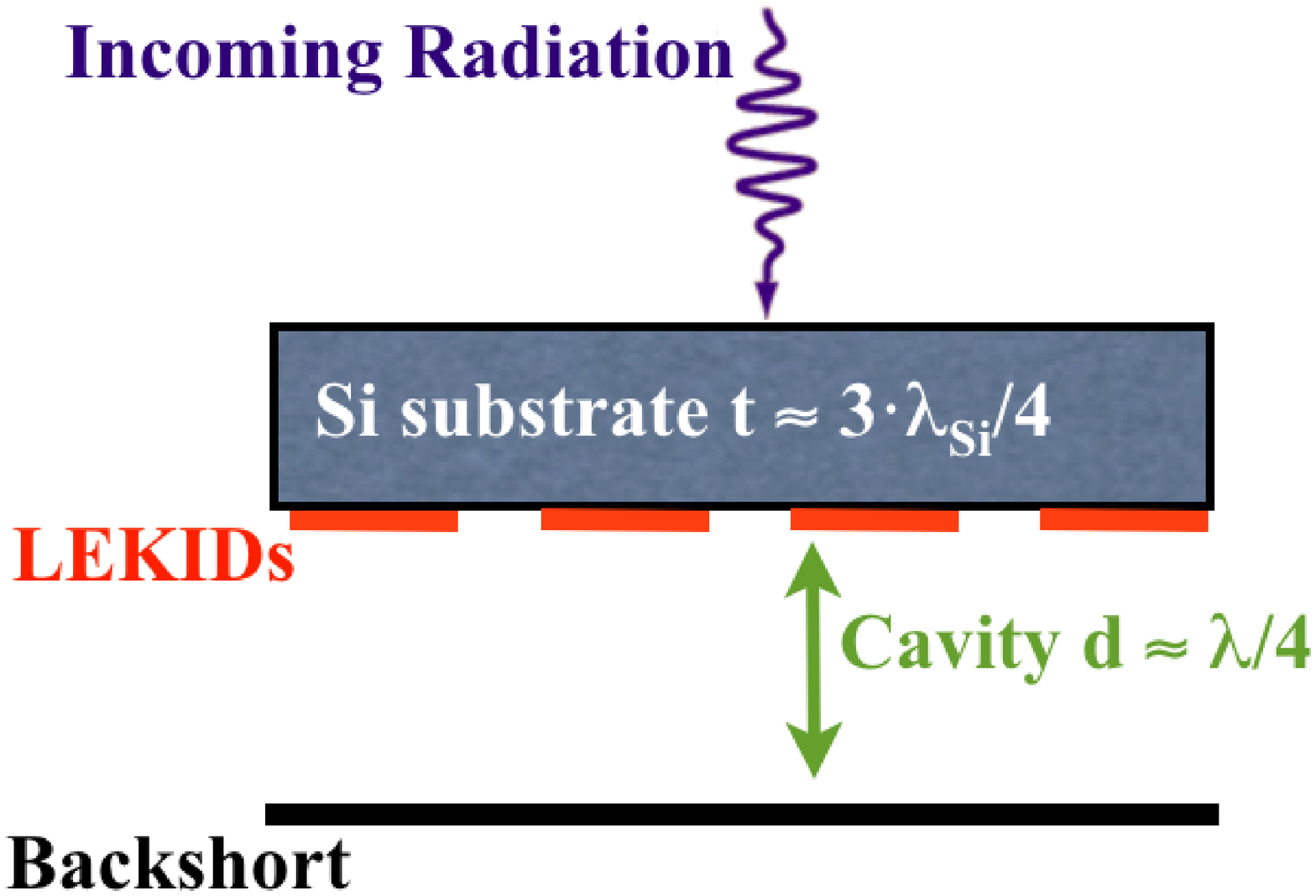}
\end{center}
  \caption{Single Ti-Al bilayer LEKID resonator design (top). The resonator is composed of a long inductive meander line and a capacitive element used also for tuning the resonant frequency. The effective sheet impedance of the meander line (seen by GHz-THz photons) is adjusted by the geometrical filling factor, for a given the film resistivity. In order to optimise 100~GHz photon absorption, the pixels are back-illuminated through the Silicon substrate. In addition we use a back-short cavity situated at an optimised distance of 750~$\mu m$
from the pixels (bottom).}
\label{fig:lekid}
\end{figure}

\begin{figure*}
\begin{center}
%\vspace{0.5cm}
\includegraphics[width=9.8cm]{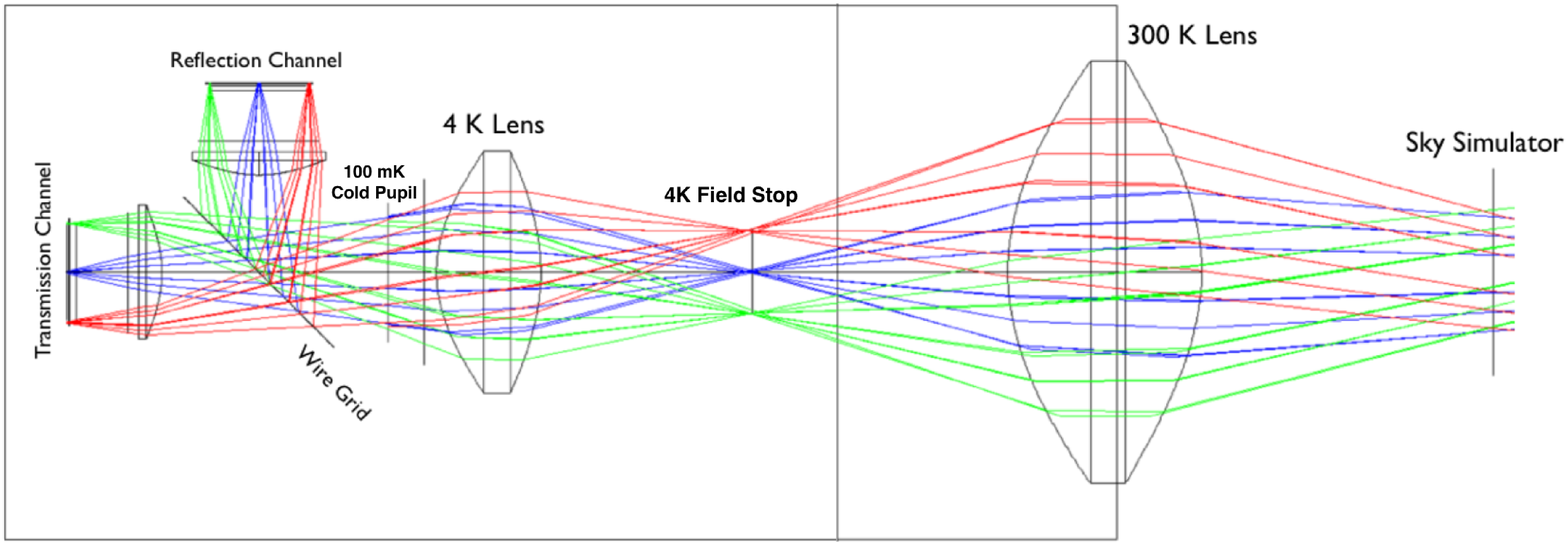}
%\vspace{2cm}
\includegraphics[width=8.5cm]{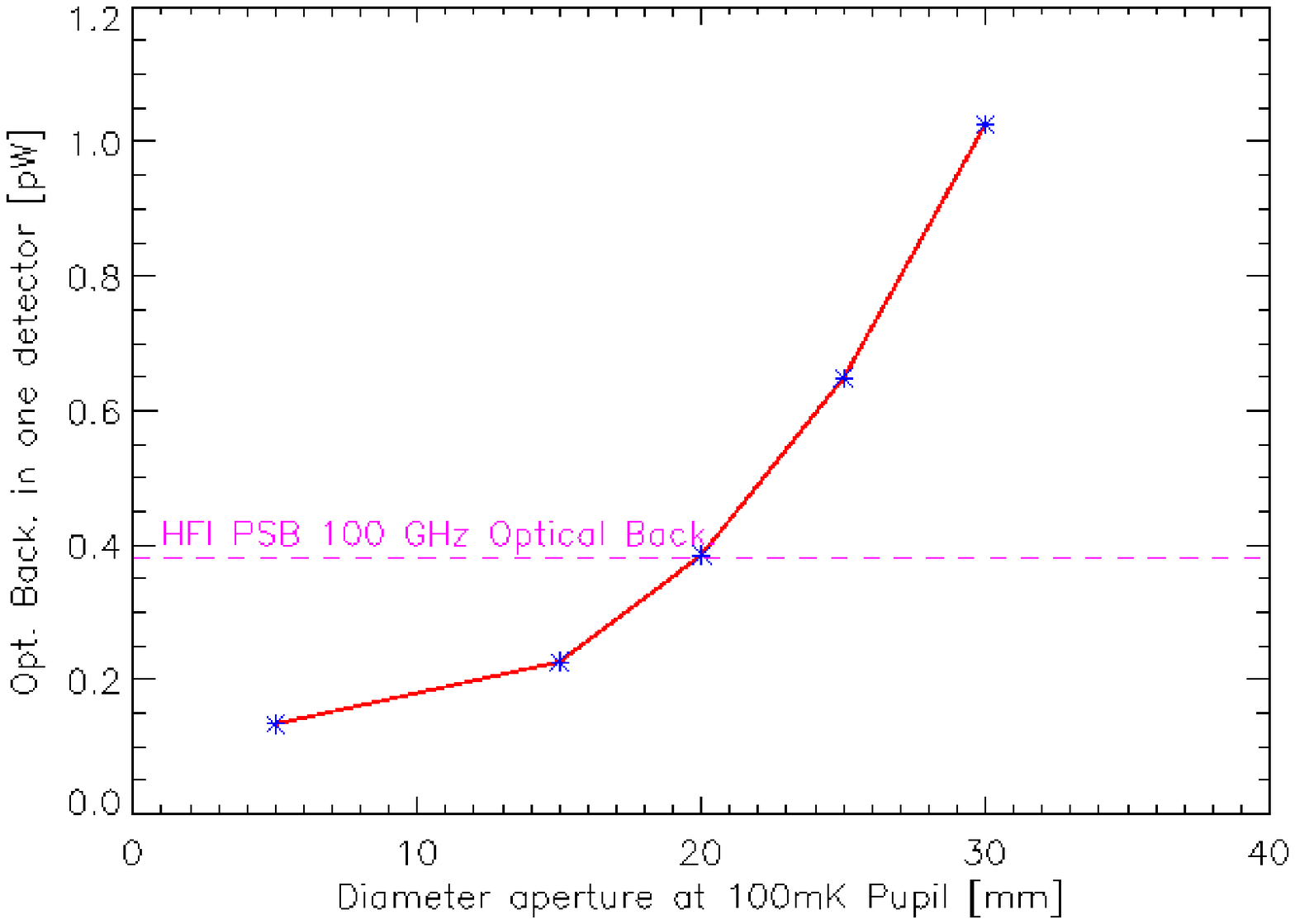}
\end{center}
  \caption{Left: Ray-tracing snapshot of the cold optics inside the cryostat. Right: optical background estimation as a function of 100~mK cold stop aperture diameter (red line). We compare this results to the optical background measured in the HFI 100~GHz channel (violet dashed line).}
\label{fig:optics}
\end{figure*}

%Adjusting the $T_c$ of the detectors is therefore important to be able to tune the absorption of photons in the suitable bandpass. 
%In this sense, bi-layers are an easy choice because the critical temperature can be tuned changing the relative thickness of the layers. If we consider Ti-Al bilayers, both materials are superconducting and Aluminium has a transition temperature higher then Titanium. In the vicinity of the transition temperature $T_c$ the proximity effect is described by the linearised set of Usadel equations \citep{usadel,2000NIMPA.444...23M}: 

%\begin{equation}\label{tc}
%\Delta S_i = ln \frac{T_c}{T_{cS_i}}+2\pi T_c \sum_{\omega_n > 0} \frac{\Delta_{S_i}-\Phi_{S_i}G_{S_i}}{\omega_n} =0
%\end{equation}

%Where $\Phi_{S_i}$ and $G_{S_i}$are the normal and anomalous Green's functions, $\Delta S_i $ is the order parameter and $\omega_n$ is the Matsubara frequency. In order to calculate $T_c$ in the general case, we have to solve the set of Eq \ref{tc} numerically. $T_c$ is defined as the maximum temperature for which non trivial solutions for the pair potentials $\Delta S_1$ and $\Delta S_2$ exist. 
%For Ti-Al bi-layers films adopted for this article, the expected $T_c$ calculation gives a solution equal to .... \NOTE{ask to Helene for this.}

\subsection{Optical Responsivity}

The optical responsivity for a LEKID represents the relation between a variation of the incident optical power and the shift in the resonant frequency of the detector ($\Re=\frac{d f}{dW}$).
This measured variation is determined by the change in internal inductance of the film with a change in Cooper pair density induced by incoming photons.

For the LEKID, all the properties directly related to the calculation of the responsivity affect each other so that it is not possible to derive a general single analytical formula. For a detailed study of the parameters that determine the responsivity of a LEKID we refer to \cite{doyle1} and \cite{doyle2}.

\subsection{Noise}\label{sec:noise}

The detector Noise Equivalent Power (NEP), in $W/\sqrt{Hz}$, is defined as the optical signal that is equal to the noise in a 1~Hz amplifier bandwidth at the output. This quantity takes into account the response and the spectral noise density Sn(f) of the detector (in $Hz/ \sqrt{Hz}$). All the uncorrelated sources of noise add quadratically:

\begin{equation}\label{nep}
NEP^2_{tot} = \sum_{i=1}^N NEP_{det(i)}^2+\sum_{i=1}^N NEP_{el(i)}^2 + NEP^2_{phot} 
\end{equation}
The detectors noise $NEP_{det}$ may come from random variations of the effective dielectric constant or fluctuations of the Cooper pair density due to generation-recombination noise\footnote{In the case of NIKA Aluminum LEKID, the contribution of the detector noise has been measured to be lower than the other sources of noise and relatively flat in frequency}. $NEP_{el}$ is the noise associated to the full readout electronics chain (cold and warm). 
Finally $NEP_{phot}$ represents the source of noise related to the photon noise which comes from the fluctuations of the incident radiation due to the Bose-Einstein distribution of the photon emission. This noise corresponds to the ultimate limitation in sensitivity of any instrument as it does not depend on detector performance and readout electronics. Considering a satellite orbiting around the second Lagrange point of the Earth-Sun system, the in-space photon noise level per pixel at 100~GHz is equal to about $0.5\cdot10^{-17}$~$W/\sqrt{Hz}$. 

As a guideline, for future generation millimetre satellite as well as for Planck HFI,  the goal NEP per pixel has been defined:

\begin{equation}
NEP_{GOAL} \leq 2 \cdot NEP_{phot}
\end{equation}

When this condition is satisfied we refer to photon noise dominated detectors. We keep this definition as a reference to compare the results obtained in this paper. 

%\NOTE{Distinct advantages of KID are the intrinsic frequency-division-multiplexing and the fast response
%time, not limited by thermal constrains}

\subsection{LEKID module design, fabrication and assembly}
For the first generation Ti-Al arrays, we adopted single polarisation LEKID designed originally for Aluminium films (Fig. \ref{fig:lekid}). The first prototype of Ti-Al arrays comprises 25 pixels. 
The LEKID holder allows back-illumination through the Silicon substrate. This method works better than  directly illuminating the LEKID as high permittivity silicon substrate has a far lower impedance than that of free space. In front of the LEKID, a superconducting lid acts as a $\lambda/4$ back-short, optimising the absorption in the frequency band of interest. A schematic of this arrangement is shown in Fig. \ref{fig:lekid}

Each pixel is a resonator composed by a meander inductor and an interdigitated capacitor.  The 4~$\mu m$ wide lines of the meander are the smallest features in the current design. The 25 pixels are inductively coupled to a coplanar wave guide (CPW) $50\Omega$ line.
The fabrication starts with a high resistivity ($>$~6000~$\Omega \cdot cm$), 525~$\mu m$ thick silicon [111] mono-crystalline wafer. The native silicon oxide is etched with an HF:$H_2O$ 5\% solution, then rinsed in $H_2O$. This step passivates the Si bonds on the surface, replacing O with H. The wafer is then put under vacuum within the next 30 minutes, to prevent re-oxidation. The bilayer is coated \emph{in situ} by e-beam evaporation under a $\approx3e-8~mb$ vacuum. A 10~nm thick titanium film is first deposited, followed directly by a 25~nm thick aluminium film. This ensures that there is no oxidation layer at the interface between titanium and aluminium, and grants us a maximal proximity effect between the two layers.
The next step is a standard photo-lithographic process, with a positive resin (AZ1512Hs). The etching is made in two phases, first with an Al-etch dip then a dilute 0.1~\% HF solution to etch Ti. After dicing, the array is mounted in a dedicated holder, and the CPW is wire-bonded to the launcher with an Al-Si wire.

\subsection{Readout system}
We used a Vector Network Analyzer (VNA) to measure the LEKID responsivity and the NIKEL electronics (described in \cite{bourrion2012}) to measure the spectral response and characterise the noise. The NIKEL electronics has been successfully used during several NIKA observing campaigns \citep{catalano, adam}. We briefly describe the readout concept: we generate a frequency comb which is up converted by mixing with a local oscillator carrier at the appropriate frequency, corresponding to the best estimate resonant frequency for each pixel. On the output line, the signal is boosted by a cryogenic low noise amplifier, then is down-converted to the base band and acquired by an Analog to Digital Converter (ADC). Finally, the useful signal is computed on-board with an FPGA: each output tone is compared to a copy of the input tones kept as a reference, so that for each pixel we extract the in phase (I) and quadrature (Q) components of the signal.

The knowledge of the I and Q components alone does not give full access to the change in resonance frequency due to an incoming optical signal. To do so we also modulate the frequency on the local oscillator (by few kHz) synchronously to the FPGA in order to recover the differential values dI, dQ in addition to I and Q components. More details of this technique are presented in \cite{Calvo2013}.

\begin{table}[b!]
\begin{center}
\begin{tabular}{cc}
\hline
\hline
 & Ti-Al Film  \\
\hline \hline
Valid Pixels [\#] &  18 \\
Pixel size [mm] &  2.3 \\
Film Thickness [$n m$] & 10-25\\
Polarised Sensitive Detectors &   yes \\
Angular size ($F \lambda$) &   0.75 \\
Overall optical efficiency [\%] &  30 \\
Total background [pW] &     0.4 \\
%$NEP_{GOAL}$ [aW/$\sqrt{Hz}$] & 10 \\
\hline \hline
\end{tabular}
\end{center}
\caption{Characteristics of the instrumental setup used for the measurements. The total background is calculated per pixel using a realistic optical simulation.}
\label{tab1}
\end{table}

\section{Optical Background Control}\label{opti}

In order to characterise the optical response of the Ti-Al array we use a testing device called \emph{Sky Simulator} (hereafter SS). It was originally built in order to mimic the typical ground optical background for the NIKA instrument at the IRAM 30~m telescope in Pico Veleta \citep{monfardini}. The working principle is based on cooling down a black disk of 25~cm diameter by using a single-stage pulse-tube refrigerator. The SS temperature can be controlled between 50~K and 300~K. This allows us to efficiently estimate the optical detector response and the NEP. The spectral response of the detectors is derived from a classic Martin-Puplett Interferometer (hereafter MPI) based on a blackbody source modulated between 300~K and 77~K and fully polarised by a wire-grid. 

The optical coupling between the SS (or the MPI) and the detectors is ensured by cold refractive optics.  
The optical system consists of a 300~K window lens, a field stop plus a lens at the 4~K stage. A 60~mm aperture stop and a lens are placed at the 100~mK stage in front of the back-illuminated LEKID arrays. Finally, in order to acquire data simultaneously with two independent RF channels a wire-grid polariser acts as beam splitter for the two arrays. The left panel of Fig~\ref{fig:optics} shows a 2-D layer snapshot from a ray-tracing simulation used to optimise the optical system.

The pixel throughput\footnote{The throughput is the name of the optical invariant that is used for the product of the pupil area and the solid angle subtended at this pupil.} is derived from the sky-simulator diameter, the fraction of un-vignetted pupil defined by the cold field stop, and the receiver pixel size with respect to diffraction pattern. We obtain:

$$
A\Omega_{pixel}  = (\pi/4)  (u \lambda)^2 
$$
where $u$ is the pixel angular size in units of $F \lambda$ ($\lambda$ = wavelength and $F$= f-number or relative aperture), which is equal to 0.75. %This means that each pixel receives about 33~\% of the incoming power.  

The spectral band is defined by a series of low-pass metal mesh filters, placed at different cryogenic stages in order to minimise the thermal loading on the detectors. A last 3.7~$cm^{-1}$ low pass filter is mounted in front of the LEKID array at 100~mK. The final bandpass is determined by this filter at high frequencies and by the superconducting gap cut-off at low frequencies. 

We performed an optical simulation of the system accounting for the absorption, reflection and emission of the polyethylene lenses and the diffracted beam due to the cold aperture stop at 100~mK to estimate the optical background on the focal plane.
The results of this simulation have been validated by comparing the optical background on NIKA Aluminium arrays for laboratory tests to the one measured at the 30~m IRAM telescope. We estimate a level of uncertainties of about 50~\%.
% have a 50\% precision level. 
In order to re-scale the total optical background on each pixel to the typical in-flight optical background at 100~GHz, we reduced the cold aperture from 60 to 20~mm obtaining an incoming optical power of about 0.4~pW (see Fig~\ref{fig:optics}). 

\begin{figure*}
\begin{center}
\includegraphics[width=9cm]{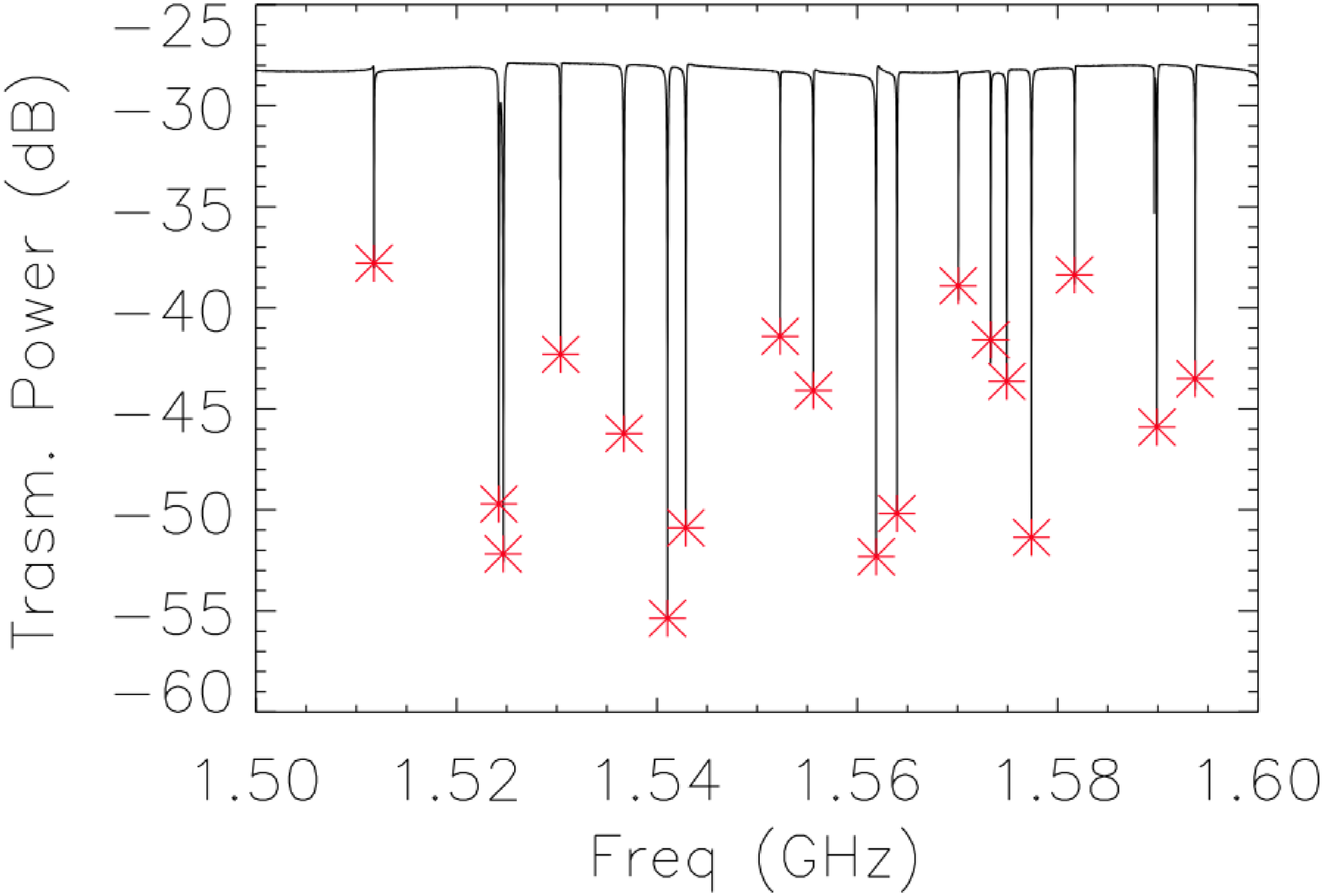}
\includegraphics[width=9.2cm]{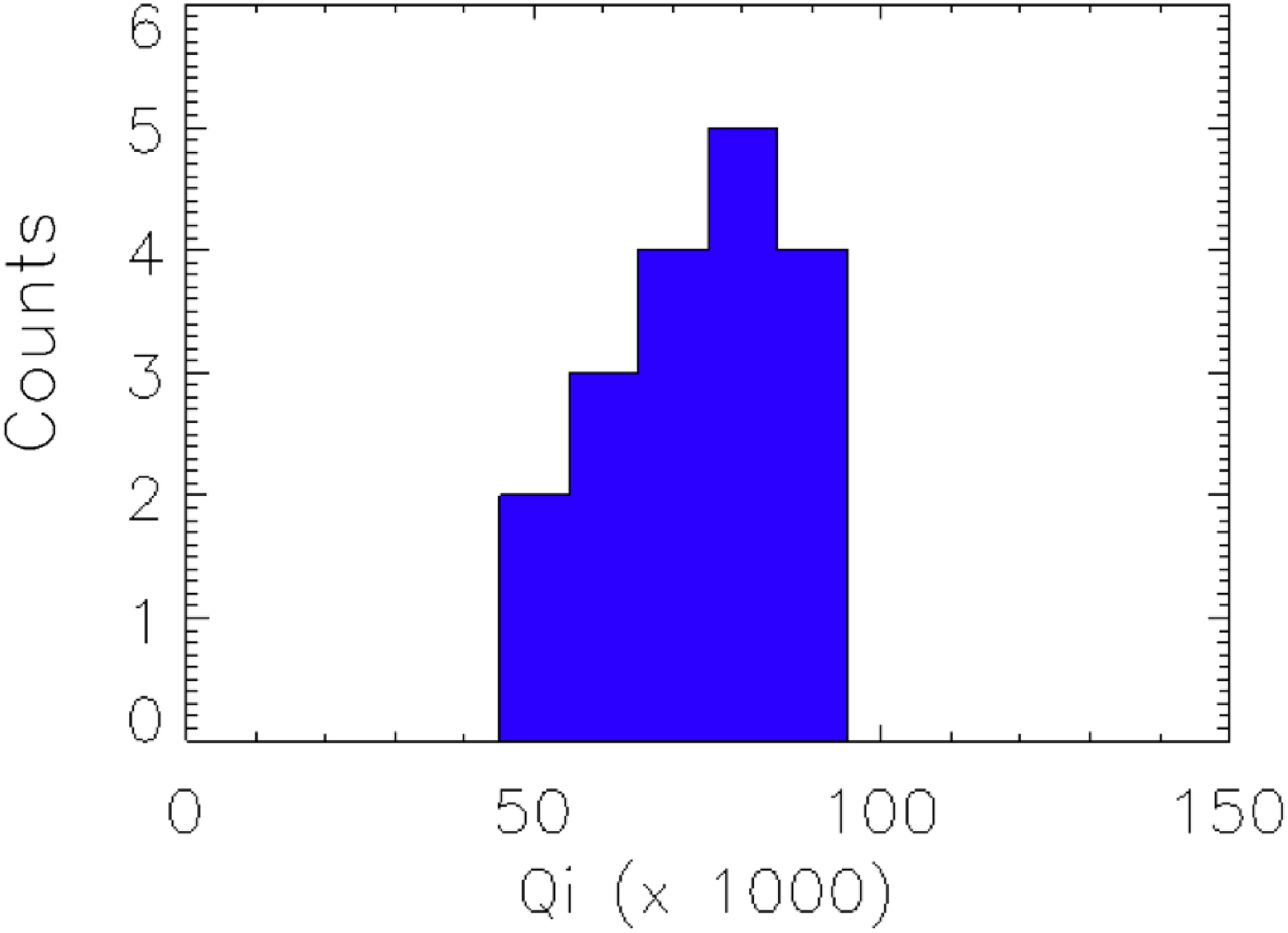}
\end{center}
  \caption{Left: frequency sweep of the first generation of 25 pixels Al-Ti array. Right: internal quality factor distribution.}
\label{fig:elect}
\end{figure*}

\section{Performance testing}\label{perf}
For a complete understanding of the detectors performance and to characterise the single-pixel sensitivity, we have performed laboratory measurements during the last year. 
The experimental tool is optimised to work under low optical background representative of the in-space sky emission at 100~GHz. The characterisation of the optical responsivity and noise is performed using as source the SS; the spectral response is measured with the MPI. The LEKID array is cooled at a base temperature of 100~mK thanks to a closed-cycle $^3$He - $^4$He dilution cryostat designed for optical measurements. The cryostat also hosts two independent RF channels, each one equipped with a cold low-noise amplifier optimised to work at the detectors frequencies.
As KIDs detectors are sensitive to the Earth magnetic fields and to those induced by the instrumentation present in the laboratory, two magnetic shields have been added to reduce this noise source: a mu-metal screen at the 300~K stage and a superconducting lead screen on the 1~K stage. We summarise in Tab~\ref{tab1} the main characteristics of the experimental setup. 
%In the next subsections we describe the results obtained for the bi-layer that presented the best performances.

\subsection{Electrical Properties}

In order to identify the LEKID resonances we performed a frequency sweep with the array connected to the VNA. The result is shown in Fig~\ref{fig:elect}. We clearly see that the feed-line is not affected by the presence of standing waves supported by the slot-line modes. This is possible thanks to the bondings added between the substrate and the holder and across the feed-line. In consequence, the depth of the different resonances (i.e. the internal quality factor) is quite uniform. The pixels resonate between 1.5 and 1.6~GHz and present homogeneous internal quality factors of $8\cdot 10^4$ (see Fig~\ref{fig:elect}). Thus, each resonance occupies a bandwidth of about 20-30~kHz for an optical background of about 0.4~pW.

The critical temperature is measured with the VNA connected to the array and warming up slowly the 100 mK stage up to the superconducting transition. The transition is observed at  $900 \pm 25$~mK in good agreement with expectations (see Sec. \ref{tc}). Using Eq. \ref{gap}, we expect a cut-off due to the superconducting gap at 65~GHz. %This cut-off is not supposed to be very steep as a consequence of transition shallowness. 

The normal-state sheet resistance of the Ti-Al bilayer has been measured around $R_s=0.5$~Ohm/square at 1~K. This value is related to the measured critical temperature and the surface inductance (estimated at 1~pH/sq as expected for the thickness considered) through the relation \citep{leduc}: 

$$
L_s=\frac{\hbar R_s}{\pi \Delta} 
$$
where $\Delta$ is the superconducting gap. The three electrical values ($\Delta$, $L_s$ and $R_s$), measured or estimated independently, are in agreement within 10\% according to the cited relation.

\begin{figure}[b!]
\begin{center}
\includegraphics[width=9cm]{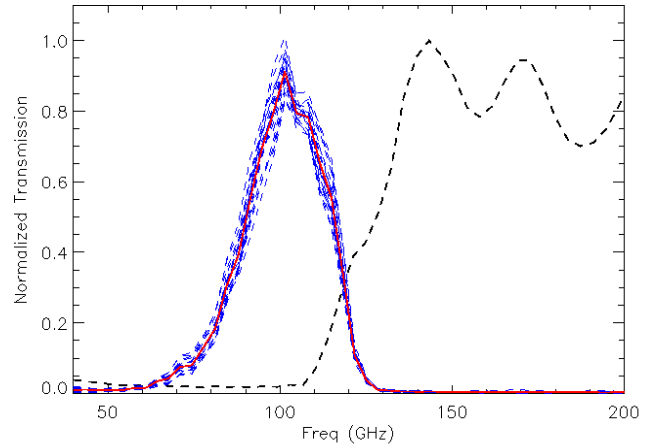}
\end{center}
  \caption{Individual detector (blue dashed lines) and averaged (red line) normalised Ti-Al spectral response ($\Delta S_i / max(\Delta S_i)$ where $\Delta S_i$ is the response of the $i-th$ pixel). The dispersion level between pixels is of the order of few per cent. For illustration, the spectral response of the Ti-Al array is compared to the one of a pure 18~nm thick Aluminium LEKID array with a $T_c$ of roughly 1.4~K (dashed black line).}
\label{fig:spectrum}
\end{figure}

\begin{figure*}
\begin{center}
\includegraphics[width=8.6cm]{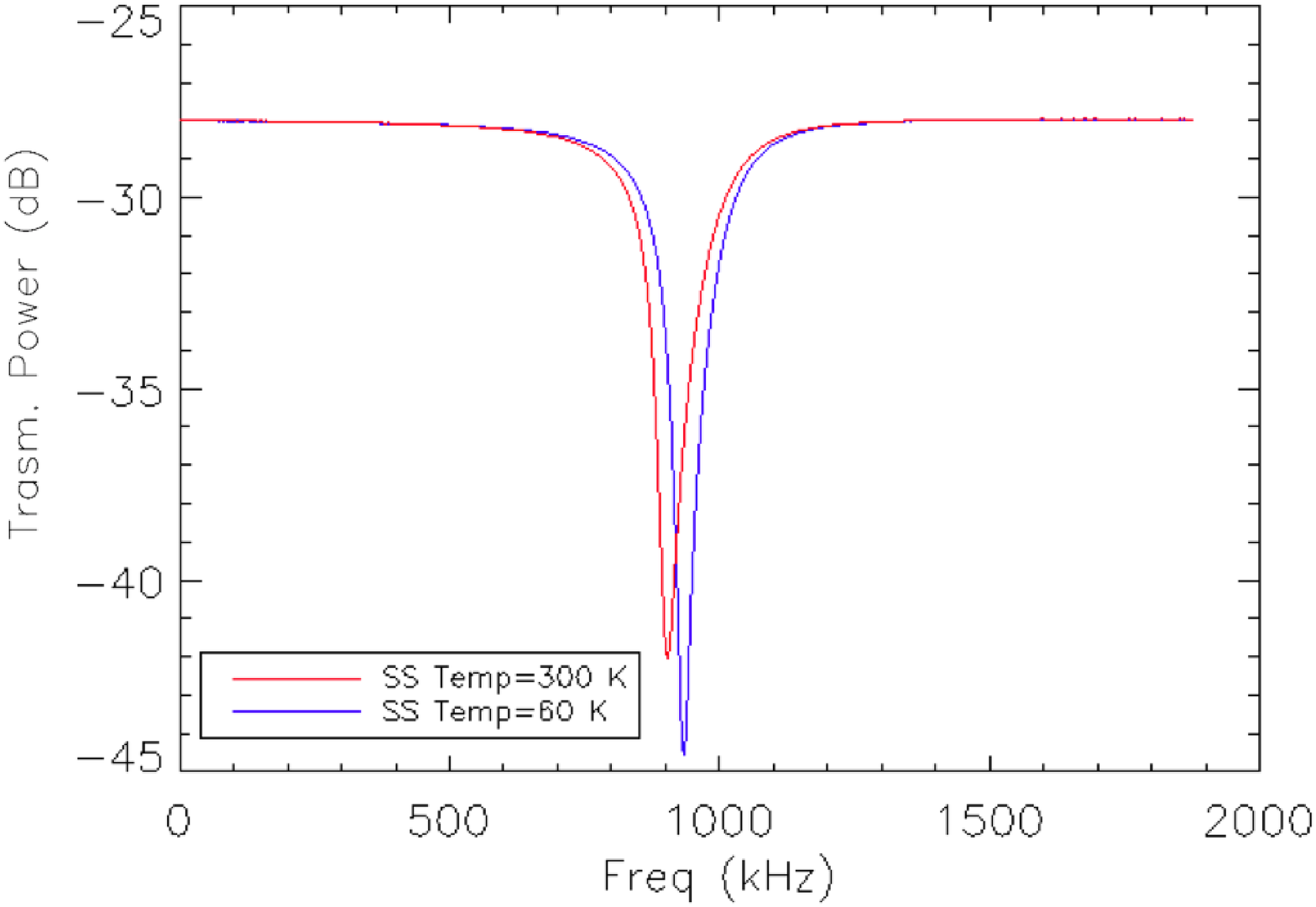}
\includegraphics[width=9cm]{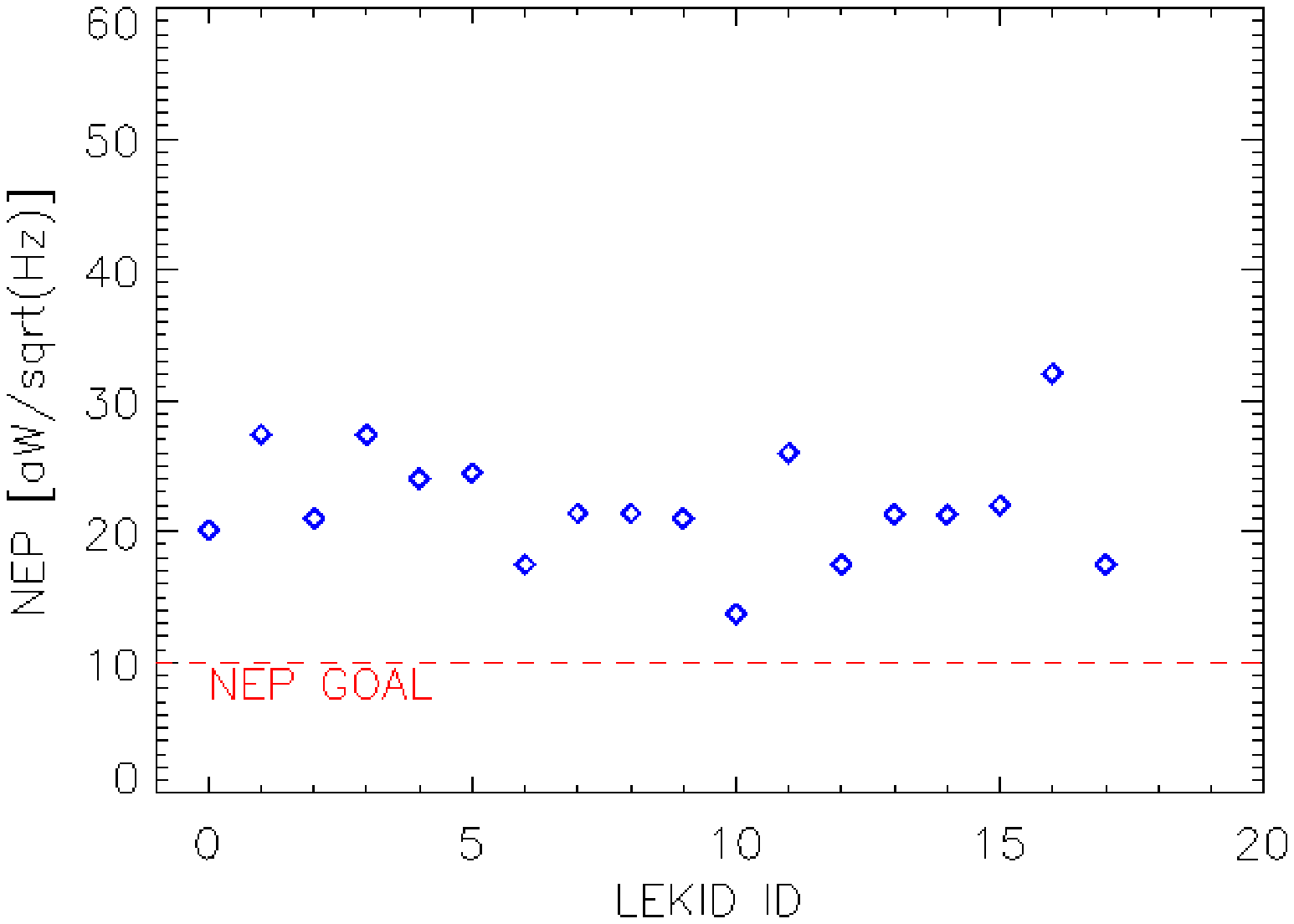}
\end{center}
  \caption{Left: resonant frequency shift due to the variation of the SS temperature from 80~K to 300~K. Right: individual pixel sensitivities (blue diamonds) measured during tests compared to the reference goal (dashed red line).}
\label{fig:shift}
\end{figure*}

\subsection{Spectral Transmission}

The gap of the Ti-Al bilayer has been measured from the absorption
spectra taken in the lab using the MPI. The results are shown in Fig~\ref{fig:spectrum}. We observe that below roughly 65~GHz a very low level of radiation is transmitted (less than 5~\%). This means that the resultant energy gap for the Ti-Al films is $\approx 135~\mu$eV, in agreement with the critical temperature. The corresponding spectrum bandwidth\footnote{the bandwidth is defined as $\Delta \nu / \nu = \frac{FWHM}{\nu_0}$ where the $FWHM$ is the full width at half maximum of the spectrum and $\nu_0$ is the central frequency.} $\Delta \nu / \nu$ has been measured equal to 0.28.
%If we define the bandwidth as $Bandwidth = \frac{FWHM}{\nu_0}$ where the $FWHM$ is full width at half maximum of the spectrum and $\nu_0$ is its central frequency, from the Fig~\ref{fig:spectrum} we measure a bandwidth of defined as :

\subsection{Optical Response and Noise Equivalent Power}

The optical responsivity of the pixels has been measured using the VNA. We perform frequency sweep to measure the LEKID transfer function for various SS background temperatures. In Fig~\ref{fig:shift} we present the shift of the resonant frequency for one pixel when the SS temperature is changed from 80 to 300~K. This frequency shift averaged across all the pixels correspond to about 27~kHz. Using the optical model we estimate the corresponding variation in optical power per pixel to about 0.3~pW. Therefore, the averaged responsivity for the array is $\Re$ = 90~$kHz/pW$.   

The spectral noise density $Sn(f)$ (in $Hz/\sqrt{Hz}$), is calculated at a fixed SS temperature of 80~K using the NIKEL electronics. 
Correlated electronic noise is removed by substracting a common mode. This is obtained by averaging the Time-Ordered-Data (TOD) of all detectors in the array. The resulting template is fitted linearly to the TOD of each detectors. The best-fit is then subtracted from the detectors TODs. After de-correlation, the spectral noise density is flat in a band between 1 and 10~Hz and equal to about $Sn(f) = 1-3$~$Hz/\sqrt{Hz}$. We can easily compute the NEP as:

$$
NEP = \frac{Sn(f)}{\Re}
$$
The right panel of Fig~\ref{fig:shift} shows the NEP of individual pixels. The best pixel has a NEP equal to  $1.4$ $10^{-17}$~$W/Hz^{0.5}$. The averaged NEP over all the array is $2.2$ $10^{-17}$~$W/Hz^{0.5}$ which is about twice the goal. 
%The NEP is presented in Fig~\ref{fig:shift} for all individual pixels and is compared to the reference goal sensitivity defined in Sect.~\ref{sec:noise}. 

\section{Conclusion}\label{concl} 

%We have produced and tested LEKID sensitive in the band between 80-120~GHz based on multi-layers superconductors films. The main results are:
We have produced and tested high quality LEKID detectors based on multi-layers superconductors films. They have proved to be sensitive in the frequency range from 80 to 120~GHz.
%In particular we find:

\begin{itemize}

\item We achieved internal quality factors exceeding $5 \cdot 10^4$ under a typical space-like optical background of 0.4~pW, and we control the transition temperature as expected from calculations. 

\item The spectral response is in agreement with the design. It peaks at 100~GHz with a 28 \% bandwidth.

\item The Noise Equivalent Power is pretty uniform across the array; the best pixels approach the reference NEP goal. On average the NEP is of the order of twice the goal. 

\end{itemize}

Notice that the LEKID design adopted for this study was originally developed for thin (less than 20 nm) Aluminium films, and ground-based typical optical backgrounds. The sensitivity could thus be further improved by optimising, for example, the films, back-short and substrate thicknesses, the resonator coupling to the RF feed-line and the meander geometry. Building on these promising results, larger arrays (hundreds pixels) will be developed in order to investigate and mitigate, as we did already for Aluminium films, systematic effects like for example crosstalk between pixels and homogeneity over the array.

%The sensitivity could be improved by optimising the preliminary LEKID design, for example the films, backshort and substrate thicknesses, the resonator coupling and meander geometry. 
%After these promising results, we are planning to produce larger arrays (hundreds pixels) in order to optimise other systematics errors like crosstalk between pixels and homogeneity over the array. 

\acknowledgement{The engineers more involved in the experimental setup development are: Gregory Garde, Henri Rodenas, Jean-Paul Leggeri, Maurice Grollier, Guillaume Bres, Christophe Vescovi, Jean-Pierre Scordilis, Eric Perbet. We acknowledge more in general the crucial contributions of the whole Cryogenics and Electronics groups at Institut N\'eel and LPSC. The arrays described in this paper have been produced at the CEA Saclay and PTA Grenoble microfabrication facilities. This work has been supported as part of a collaborative project, SPACEKIDS, funded via grant 313320 provided by the European Commission under Theme SPA.2012.2.2-01 of Framework Programme 7.}

\bibliographystyle{aa}
\bibliography{cata}

\end{document}